\documentclass[aps,prl,floatfix,twocolumn,amsmath,amssymb]{revtex4-1}
\usepackage{amsthm,color,latexsym,array,epstopdf}
\usepackage[hyperfootnotes=true,breaklinks]{hyperref}
\usepackage{graphicx}
\usepackage{grffile}

\usepackage{feynmp-auto}
\usepackage{bbm,slashed,tikz}
\definecolor{darkred}{rgb}{0.5,0.0,0.0}
\definecolor{darkblue}{rgb}{0.0,0.0,0.9}
\definecolor{darkerblue}{rgb}{0.0,0.0,0.5}
\definecolor{darkgreen}{rgb}{0.0,0.5,0.0}
\definecolor{black}{rgb}{0.0,0.0,0.0}
\definecolor{brown}{rgb}{0.6,0.4,0.2}

\def\be{\begin{equation}}
\def\ee{\end{equation}}
\def\cL{\mathcal{L}}
\def\cO{\mathcal{O}}
\def\cN{\mathcal{N}}

\def\cM{\mathcal{M}}
\def\Has{H_{\text{as}}}
\def\Hsc{\Has}
\def\as{\text{as}}
\def\msbar{\overline{\text{MS}}}

\def\LPeq{\cong}

\begin{document}

\begin{fmffile}{feyngraph}
\unitlength = 1mm

\title{A Finite \textit{S}-Matrix}

\author{Holmfridur Hannesdottir}
\email{holmfridur\_hannesdottir@g.harvard.edu}
\affiliation{Department of Physics, Harvard University, Cambridge, MA 02138}
\author{Matthew D.~Schwartz}
\email{schwartz@g.harvard.edu}
\affiliation{Department of Physics, Harvard University, Cambridge, MA 02138}

\begin{abstract}
When massless particles are involved, the traditional scattering matrix
($S$-matrix) does not exist: it has no rigorous non-perturbative definition and
has infrared divergences in its perturbative expansion. The problem can be
traced to the impossibility of isolating single-particle states at asymptotic
times. On the other hand, the troublesome non-separable interactions are often
universal: in gauge theories they factorize so that the asymptotic evolution
is independent of the hard scattering. Exploiting this factorization property,
we show how a finite ``hard'' $S$-matrix, $S_H$, can be defined by replacing the
free Hamiltonian with a soft-collinear asymptotic Hamiltonian. The elements of
$S_H$ are gauge invariant and infrared finite, and exist even in conformal field theories. 
One can interpret elements
of $S_H$ alternatively 1) as elements of the traditional $S$-matrix between dressed states,
2) as Wilson coefficients, or 3) as remainder functions. 
These multiple interpretations provide different insights into the rich structure of $S_H$.
% For example $S_H$ exhibits symmetries, such as dual conformal invariance, that are not symmetries of the traditional infrared-divergent $S$-matrix.
 \end{abstract}

\maketitle

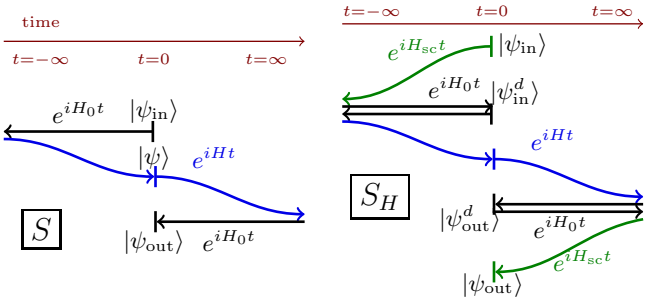
\begin{figure}[t]
{{
\begin{tikzpicture}
\draw [|->,line width=1,black] (0,0.1) to (-2,0.1);
\draw [<-,line width=1,darkblue] (0,-0.5) to [out = 180, in = 0] (-2,0);
\draw [|->,line width=1,darkblue] (0,-0.5) to [out = 0, in = 180] (2,-1);
\draw [|<-,line width=1,black] (0,-1-0.1) to (2,-1-0.1);
\node[above] at (0,0.1) {$|\psi_{\text{in}}\rangle$};
\node[above,black] at (-1,0.1) {$e^{i H_0 t}$};
\node[] at (0,-0.25) {$|\psi\rangle$};
\node[above,darkblue] at (0.8,-0.5) {$e^{i H t}$};
\node[below] at (0,-1.1) {$|\psi_{\text{out}}\rangle$};
\node[below,black] at (1,-1.1) {$e^{i H_0 t}$};
\node[darkred] at (-1.5,1) {${}^{t=-\infty}$};
\node[darkred] at (0,1)  {${}^{t=0}$};
\node[darkred] at (1.5,1)  {${}^{t=\infty}$};
\draw [->, line width=0.5,darkred] (-2,1.3) to (2,1.3);
\node[darkred] at (-1.5,1.5)  {${}^{\text{time}}$};
\node[] at (-1.5,-1.2) {\large $\boxed  S$};
\node[] at (-1.5,-2.1) {};
\end{tikzpicture}
\hspace{5pt}
\begin{tikzpicture}
\draw [|->,line width=1,darkgreen] (0,1) to [out = 180, in = 0] (-2,0.3);
\draw [<-,line width=1,black] (0,0.2) to (-2,0.2);
\draw [|->,line width=1,black] (0,0.1) to (-2,0.1);
\draw [<-,line width=1,darkblue] (0,-0.5) to [out = 180, in = 0] (-2,0);
\draw [|->,line width=1,darkblue] (0,-0.5) to [out = 0, in = 180] (2,-1);
\draw [|<-,line width=1,black] (0,-1.1) to (2,-1.1);
\draw [->,line width=1,black] (0,-1.2) to (2,-1.2);
\draw [|<-,line width=1,darkgreen] (0,-2) to [out = 0, in = 190] (2,-1.3);
%\node[] at (0,-0.25) {$|\psi\rangle$};
\node[above] at (0.4,0.75) {$|\psi_{\text{in}}\rangle$};
\node[above] at (0.3,0.1) {$|\psi_\text{in}^d\rangle$};
\node[below] at (0,-1.9) {$|\psi_{\text{out}}\rangle$};
\node[below] at (-0.3,-1) {$|\psi_\text{out}^d\rangle$};
\node[above,darkgreen] at (-1,0.7) {$e^{i H_\text{sc} t}$};
\node[above,black] at (-0.5,0.2) {$e^{i H_0 t}$};
\node[above,darkblue] at (0.8,-0.5) {$e^{i H t}$};
\node[below,black] at (0.9,-1.15) {$e^{i H_0 t}$};
\node[below,darkgreen] at (1.2,-1.6) {$e^{i H_\text{sc} t}$};
\draw [->, line width=0.5,darkred] (-2,1.3) to (2,1.3);
%\node at (-1.5,1.2)  {${}^{\text{time}}$};
\node[darkred] at (-1.6,1.4) {${}^{t=-\infty}$};
\node[darkred] at (0,1.4)  {${}^{t=0}$};
\node[darkred] at (1.6,1.4)  {${}^{t=\infty}$};
\node[] at (-1.5,-1)  {\large ${\boxed {S_H}}$};
%\node at (-1.5,2) {${}^{t=-\infty}$};
%\node at (0,2)  {${}^{t=0}$};
%\node at (1.5,2)  {${}^{t=\infty}$};
\end{tikzpicture}
}}
%\vspace{5mm}
  \caption{(Left) The traditional $S$-matrix is
  computed from Fock states evolved using $H_0$ and $H$.
   (Right) The hard $S$-matrix is computed either using Fock states evolved with $\Has$ and $H$  or using dressed states evolved with $H_0$ and $H$.
\label{fig:SH}}
\end{figure}

One of the most fundamental objects in high energy physics is the scattering- or
$S$-matrix. Not only is it a bridge between a definition of a quantum theory and data from particle colliders, but the study of the $S$-matrix itself
has led to deep insights into the mathematical and physical pillars of quantum field
theory itself.
The idea
behind the $S$-matrix is that it gives the amplitude for a set of particles in
an ``in'' state $| \psi_{\text{in}} \rangle$ at $t = - \infty$ to turn into a
different set of particles in an ``out'' state $\langle \psi_{\text{out}} |$
at $t = + \infty$. To go from this intuitive picture to a mathematically
rigorous definition of the $S$-matrix has proven remarkably challenging. For
example, suppose we take the in and out states to be eigenstates of the
Hamiltonian $H$ with energy $E$. Then they would evolve in time only by a phase rotation and
the $S$-matrix elements would all have the form $\lim_{t \rightarrow \infty}
e^{ -2i E t} \langle \psi_{\text{out}} | \psi_{\text{in}} \rangle$. Such an $S$-matrix would be both
ill-defined (because of the limit) and trivial (because of the projection). In
non-relativistic quantum mechanics, one avoids this infinitely oscillating
phase by subtracting from $H$ the free Hamiltonian $H_0 = \frac{\vec{p}\,^2}{2
m}$. More precisely, one looks for states $| \psi \rangle$ which, when evolved
with the full Hamiltonian, agree with in and out states evolved with the free
Hamiltonian: $e^{-i H t} | \psi \rangle \rightarrow e^{- i H_0 t} |
\psi_{\text{in}} \rangle$ as $t \rightarrow - \infty$ and $e^{-i H t} | \psi
\rangle \rightarrow e^{- i H_0 t} | \psi_{\text{out}} \rangle$ as $t
\rightarrow + \infty$. Then the projection of  in states onto  out states
is given by matrix elements $\langle \psi_{\text{out}} | S | \psi_{\text{in}}
\rangle$ of the operator $S =  \Omega_+^{\dag} \Omega_-$ where
the M\o ller operators  are  defined as $\Omega_{\pm} =
e^{i H t_\pm} e^{- i H_0 t_\pm}$, with $t_\pm$ shorthand for the
$t\to \pm \infty$ limit.
In this way, the free evolution, which is responsible for the
infinite phase, is removed. Note that $\lim_{t\to\pm\infty} e^{-i H t} |\psi\rangle$ is not a well-defined state,
 so the in and out states should be thought of as either Heisenberg picture states
 or as Schr\"odinger picture states at $t=0$ not at $t=\pm \infty$
(see Fig.~\ref{fig:SH}).
Defining the $S$-matrix this way  gives sensible results and a pleasing
physical picture: particles we scatter are free when not interacting. Their
freedom means they should have momentum defined by the free Hamiltonian and
the $S$-matrix encodes the effects of interactions impinging on this freedom.

In quantum field theory, a similar construction is fraught with complications.
The M\o ller operators which convert from the Heisenberg picture to the
interaction picture, do not exist as unitary operators acting on a Fock space
(Haag's theorem~\cite{Haag:1955ev}). So one must work entirely in the Heisenberg picture without
reference to $H_0$. The matching of the states at $t \rightarrow \pm \infty$
is then replaced with an asymptotic condition on the matrix elements of
fields. 
 In the Haag-Ruelle construction~\cite{Haag:1958vt,Haag:1959ozr,ruelle1962asymptotic}, a mass gap is required
 to isolate the few-particle asymptotic states as limits of carefully constructed wave packets. From there, one can
derive the LSZ reduction theorem, relating elements of the $S$-matrix to
time-ordered products of fields~\cite{Lehmann:1954rq,Collins:2019ozc}. 

While it is satisfying to know that the $S$
matrix can be rigorously defined, its existence requires a theory
with a mass gap, a unique vacuum state, and fields whose two-point functions
vanish exponentially at spacelike separation. None of these requirements hold
in any real-world theory.
The practical resolution to this impasse is to
ignore Haag's theorem, ignore that charged particles cannot be isolated and other assumptions, and simply use
the LSZ reduction theorem as if it were true, computing $S$-matrix elements
with Feynman diagrams. Although the resulting matrix elements are singular
(due to infrared divergences) as long as one combines $S$-matrix elements
computed this way into observable cross sections, the singularities will drop
out. This is guaranteed by the KLN theorem~\cite{Kinoshita:1962ur,Lee:1964is} which
says that infrared divergences will cancel when initial and final states are
summed over, or by its stronger version, that the cancellation occurs when
initial {\it{or}} final states are summed over~\cite{Frye:2018xjj}. Despite the success 
of this pragmatic approach, it remains deeply unsettling that the
underlying object we compute, the $S$-matrix, has no formal definition even in
QED.

There has been intermittent progress on constructing an $S$-matrix for QED (and QCD)
over the last 50 or so years. The infrared divergence problem of the $S$-matrix can be seen already in non-relativistic scattering off a Coulomb
potential. Because of its $\frac{1}{r}$ behavior, the Coulomb potential is not
square integrable, and the asymptotic states do not exist. This complication
was observed by Dollard~\cite{dollard1971quantum}, and resolved by using a modified Hamiltonian
$H_{\text{as}} (t)$ that appends the dominant large-distance behavior of the
Coulomb interaction to the free Hamiltonian. Chung~\cite{Chung:1965zza}, independently, observed
that if instead of scattering single-particle Fock-state elements, one
scatters linear combinations of these elements, similar to coherent states
used in quantum optics (and to an early attempt by Dirac~\cite{dirac1955gauge}), finite amplitudes would result. In Chung's
construction, the IR divergent phase space integrals from cross-section calculations are moved
 into the definition of the states. 
 Faddeev and Kulish~\cite{Kulish:1970ut} subsequently redefined the $S$-matrix to include
 the dominant long-distance interactions of QED in its asymptotic Hamiltonian (similar to Dollard), and identified
 Chung's coherent states as arising during the asymptotic evolution.
Over the years, various subtleties in the
coherent-state approach to soft
singularities in QED have been explored~\cite{Kibble:1968sfb,Zwanziger:1973if,Bagan:1999jf}, and attempts have been made to
construct a finite $S$-matrix for theories like QCD with massless charged
particles and hence collinear singularities~\cite{del1989collinear,Contopanagos:1991yb,Giavarini:1987ts,Forde:2003jt}.

Remarkably, in all this literature, there are very few explicit calculations
of what a finite $S$-matrix looks like. Indeed almost all of the papers
concentrate on the singularities alone. Doing so sidesteps the challenge how
to handle finite parts of the amplitudes and precludes the possibility of
actually calculating anything physical. With an explicit prescription, you
have to contend with questions such as: what quantum numbers do the
dressed states have? They cannot have well-defined energy and momentum outside of
the singular limit, since they are superpositions of states with different
numbers of non-collinear finite-energy particles. 

The basic aspiration of much of this literature is that when there are 
long-range interactions, the $S$-matrix should be defined through asymptotic M\o ller operators $\Omega_{\pm}^{\text{as}} =
 e^{ i H t_\pm}e^{-i H_{\text{as}} t_\pm}$ with
 some kind of asymptotic Hamiltonian $H_{\text{as}}$ replacing the free Hamiltonian $H_0$.
 Despite the simple summary, working out the details and establishing a productive calculational framework 
has proved a resilient challenge.

In this paper we continue the quest for a finite $S$-matrix by folding into
the previous analysis insights from the modern understanding of scattering
amplitudes and factorization. We argue that the principle by which the
asymptotic Hamiltonian is to be defined is not that the dominant long-distance
interactions be included (which allows for $H_{\text{as}} = H$ and $S =
\mathbbm{1}$), but that the evolution of the states be independent of how they
scatter. 

In gauge theories, infrared divergences can be either soft or
collinear in origin. Both soft and collinear interactions are universal and
can be effectively separated from the remainder of the scattering process. 
Factorization has been understood from many perspectives~\cite{Collins:1988ig,Collins:1989gx,Bauer:2000ew,Beneke:2002ph,Beneke:2002ni,Bauer:2002nz,Bauer:2000yr,Feige:2013zla,Feige:2014wja}. A
precise statement of factorization can be found in~\cite{Feige:2013zla,Feige:2014wja}, 
where it is
proven that the IR divergences of any $S$-matrix in QCD are reproduced by the
product of a hard factor, collinear factors for each relevant direction, and a
single soft factor. A useful language for understanding factorization is
Soft-Collinear Effective Theory (SCET)~\cite{Bauer:2000ew,Beneke:2002ph,Beneke:2002ni,Bauer:2002nz,Bauer:2000yr,Becher:2014oda,stewart2013lectures}. The
SCET Lagrangian is
\be
  \mathcal{L}_{\text{SCET}} = - \frac{1}{4} (F_{\mu \nu}^s)^2
   + \sum_n -\frac{1}{4} (F_{\!\mu \nu}^{c, n})^2 
\ee
\vspace{-5mm}
\be
+ \sum_n
    \bar{\psi}_n^c \frac{\slashed{\bar{n}}}{2}
  \left[ i n \cdot D + i \slashed{D}_{c \perp} \frac{1}{i \bar{n} \cdot D_c} i
  \slashed{D}_{c \perp} \right] \psi_n^c + \cL_\text{Glauber} \nonumber
\ee
where $s$ and $c,n$ are soft and collinear labels respectively; these act like
quantum numbers for the fields. 
The derivation of the SCET Lagrangian and more details on the notation an be found in
the reviews~\cite{Becher:2014oda,stewart2013lectures}. The Glauber interactions 
denoted by $\cL_\text{Glauber}$ are discussed in~\cite{Rothstein:2016bsq}; when they are included,
the SCET Lagrangian can reproduce all of the IR singularities of any non-Abelian gauge theory.
 The main relevant features of the SCET Lagrangian
are that 1) there are no interactions between fields with different
collinear-direction labels (up to Glauber effects) and 2) collinear particles going in different
directions only interact through soft photons or gluons with eikonal interactions.
We define the asymptotic Hamiltonian $\Has$ as the SCET
Hamiltonian appended with the free Hamiltonians for massive particles.

In collider physics applications, one typically adds to the SCET Hamiltonian a
set of operators necessary to reproduce the hard scattering of interest. For
example, one might add $\Delta \mathcal{H} = C \bar{\psi} \gamma^{\mu} \psi$
for jet physics applications in $e^+ e^-$ collisions. Then one determines the
Wilson coefficient $C$ by choosing it such that matrix elements computed
using SCET agree with matrix elements computed in the full theory.
Importantly, the infrared divergences cancel in the difference, so that $C$ is IR-finite order-by-order in perturbation
theory. Motivated by such cancellations, we define hard M\o ller operators as
$\Omega_{\pm}^H =e^{i H t_\pm} e^{- i \Has t_\pm}$
and the {\bf{hard $S$-matrix}} as $S_H =
 \Omega^{H\dag}_+\Omega_-^H$. Because $\Has$ reproduces the
IR-divergence-generating soft and collinear limits of $H$, we expect the
hard $S$-matrix will be IR-finite.

To evaluate matrix elements of $S_H$ in perturbation theory, one
could attempt to work out Feynman rules in an interaction picture based on
$\Hsc$ instead of $H_0$. A propagator would then be a Green's
function for $\Hsc$, which has no known closed-form expression.
Alternatively,  we can write $S_H$ suggestively as (cf.~\cite{Kulish:1970ut,Contopanagos:1991yb})
\be
S_H =\Omega_+^{H\dag} \Omega^H_-  = \Omega_+^{\as}\Omega_+^{\dag} \Omega_- \Omega^{\as\dag}_- 
\ee
where $ \Omega^\as_{\pm} = e^{ i H_0 t_\pm}e^{-i \Hsc t_\pm}$. This encourages us to define
\be
| \psi_{\text{in}}^d \rangle = \Omega_-^{\as\dag} |\psi_\text{in}\rangle
\quad \text{and}\quad
| \psi_{\text{out}}^d \rangle = \Omega_+^{\as\dag} |\psi_\text{out}\rangle
\ee
as dressed in and out states. Then,
\be
\langle \psi_{\text{out}} | S_H | \psi_{\text{in}} \rangle  = \langle \psi_{\text{out}}^d | S | \psi_{\text{in}}^d \rangle
\ee
\vspace*{-20pt}

\noindent 
We will take  $|\psi_{\text{in}} \rangle$ and $| \psi_{\text{out}} \rangle$ to be eigenstates of the free momentum operator $P_0^\mu$ with
a few (finite number of) particles in them.  Thus we can think of $S_H$ as computing either projections among few-particle states with
the hard M\o ller operators or projections of dressed
states with the original $S$-matrix M\o ller operators.
For example, in the process $e^+ e^- \rightarrow Z$ in QED, $|
\psi_{\text{in}} \rangle$ would be an $e^+ e^-$ state of definite momentum and
$| \psi_{\text{in}}^d \rangle$ a superposition of $| e^+ e^- \rangle$, $| e^+
e^- \gamma \rangle$, $| e^+ e^- \gamma \gamma \rangle$, and so on.

More explicitly, we can relate $| \psi_{\text{in}}^d \rangle$ to $| \psi_{\text{in}} \rangle$ using time-ordered perturbation theory (TOPT). 
For example, if $| \psi_{\text{in}} \rangle$ is the state of
an electron with momentum $\vec{p}$, then in QED
\begin{multline}
\left|\psi_{\mathrm{in}} ^{d}\right\rangle=
\left| \bar{u}_s(p)\right\rangle
+ e \sum_{\epsilon} \int \frac{d^{d-1} k}{(2 \pi)^{d-1}}
 \\
\times  \frac{1}{2 \omega_{p}}\frac{1}{2 \omega_{k}} \frac{2\,p \cdot \epsilon}{\omega_k- \frac{\vec{p} \cdot \vec{k}}{\omega_{p}}-i \varepsilon}
\left| \bar{u}_s(p-k), \epsilon(k)\right\rangle+\cdots
\end{multline}
%\vspace*{-30pt}
%
%\noindent 
The denominator factor comes from the soft expansion of the TOPT propagator 
$(\omega_{p-k} +\omega_k -\omega_p -i\varepsilon)^{-1}$.
 Note that the states in the expansion of $|\psi_{\text{in}}^d \rangle$ have different energies. Although electric charge
and 3-momentum are conserved, energy is not as we evolve with $\Omega_-^\as$ in TOPT. 
Due to the IR-divergent integral over $\vec{k}$, dressed states do not exist (in contrast to $|\psi_\text{in}\rangle$ and $|\psi_\text{out}\rangle$), but they do provide a useful qualitative handle on scattering.

As a concrete example, we now compute $S_H$ for deep-inelastic scattering, $e^- \gamma^\star \to e^-$ in QED with massless fermions at momentum transfer
$Q = \sqrt{-q^2}$ in the Breit frame. 
At order $e^2$, the loop contribution to the  $S$-matrix element is, in $\msbar$ and $d=4-2\epsilon$ dimensions~\cite{Manohar:2003vb},
%\vspace*{-10pt}
\be
\cM_A
= 
\begin{gathered}
\begin{tikzpicture}
 \node at (0,0) {
\parbox{20mm} {
\resizebox{20mm}{!}{
     \fmfframe(0,00)(0,0){
 \begin{fmfgraph*}(40,20)
	\fmftop{R1,R2}
	\fmfbottom{L1}
	\fmf{fermion}{v1,L1}
	\fmf{fermion}{R1,v1}
	\fmf{fermion}{v2,R2}
	\fmf{fermion}{L1,v2}
	\fmf{photon,tension=0}{v1,v2}
        \fmfv{d.sh=circle, d.f=30,d.si=0.1w}{L1}
\end{fmfgraph*}
}}}};
\draw[dashed, line width = 1, darkred] (-0.7,0.5) to (-0.7,-0.7);
\draw[dashed, line width = 1, darkred] (0.7,0.5) to (0.7,-0.7);
\node[above,darkred] at (-0.7,0.4) {${}^{t=-\infty}$};
\node[above,darkred] at (0.7,0.4) {${}^{t=\infty}$};
\end{tikzpicture}
\end{gathered}
\hspace{4.2cm}
\ee
\vspace{-8mm}
\be
\nonumber
\hspace{-2mm}=\cM_0 \frac{\alpha}{4 \pi} 
\Big[\frac{1}{\epsilon_{\mathrm{UV}}}-\frac{2}{\epsilon_{\mathrm{IR}}^{2}}-\frac{2 \ln \frac{\mu^{2}}{Q^{2}}+4}{\epsilon_{\mathrm{IR}}} 
-\ln ^{2} \frac{\mu^{2}}{Q^{2}}-3 \ln \frac{\mu^{2}}{Q^{2}}-8+\frac{\pi^{2}}{6} \Big]
\ee
%\vspace*{-20pt}

\noindent
with $\cM$ defined by $S_H = \mathbbm{1} + (2\pi)^4\delta^4(q+p_1-p_2)i\cM$ and $\cM_0 = -e \bar{u}(p_2)\gamma^\mu u(p_1)$ is the tree-level amplitude.

While this $S$-matrix element is IR-divergent, there are other contributions to $S_H$ at the same order. These can be thought of as $S$-matrix elements for the $e^-\gamma$ components of $|\psi_\text{in}^d\rangle$ or $|\psi_\text{out}^d\rangle$. We can represent the new graphs as cuts through a broader graph, going from $0 \to -\infty \to \infty \to 0$. The first and last transitions go backward in time and represent the dressing and undressing of the state in the asymptotic regions. 
For example, the graph with both photon vertices coming from soft-collinear interaction in $\Hsc$ is
\vspace*{-10pt}
 \begin{multline}
 \cM_B
= 
\begin{gathered}
\begin{tikzpicture}
 \node at (0,0) {
\parbox{20mm} {
\resizebox{20mm}{!}{
     \fmfframe(0,00)(0,0){
 \begin{fmfgraph*}(40,20)
	\fmftop{R1,R2}
	\fmfbottom{L1}
	\fmf{fermion,label=$\searrow \!  p_1$,l.d=0.3,l.s=left}{R1,v1}
	\fmf{fermion,label=$p_2\! \nearrow $,l.s=left,l.d=0.3}{v2,R2}
	\fmf{photon,label=$\rightarrow k$,tension=0,l.s=left}{v1,v2}
	\fmf{fermion}{v1,L1}
	\fmf{fermion}{L1,v2}
        \fmfv{d.sh=circle, d.f=30,d.si=0.1w}{L1}
\end{fmfgraph*}
}}}};
\draw[dashed, line width = 1, darkred] (-0.3,0.5) to (-0.3,-0.7);
\draw[dashed, line width = 1, darkred] (0.3,0.5) to (0.3,-0.7);
\node[above,darkred] at (-0.4,0.4) {${}^{t=-\infty}$};
\node[above,darkred] at (0.4,0.4) {${}^{t=\infty}$};
\draw[->,line width=0.2,black] (-0.35,-0.8) to (-0.1,-0.6);
\node[below] at (-0.15,-0.57) {\tiny $q$};
\end{tikzpicture}
\end{gathered}
=  \cM_0 e^2 \mu^{4-d}
\int \frac{d^{d-1} k}{(2 \pi)^{d-1}}
\\[-7pt]
\times 
\frac{1}{2\omega_k}\frac{1}{2\omega_1}\frac{1}{2\omega_2}
\frac{8 \omega_1 \omega_2}{\omega_k - \frac{\vec p_1\cdot \vec k}{\omega_1} - i \varepsilon} 
 \frac{1}{\omega_k - \frac{\vec p_2\cdot \vec k}{\omega_2} - i \varepsilon}
\end{multline}
To derive this integrand, we have power-expanded in the soft limit as in the method-of-regions approach~\cite{Beneke:1997zp}, rather than using $\cL_\text{SCET}$ directly. 
Although energy is not conserved in the asymptotic regions, the central region gives $\delta(\omega_k + \omega_{p_1 - k} - \omega_k - \omega_{p_2 - k})\LPeq \delta(\omega_1-\omega_2)$ which is factored out in the definition of $\cM$. 

This integral is scaleless and vanishes. Although we cannot  easily separate all the UV and IR poles,
 the double soft/collinear pole in this amplitude is
 \be
\cM_B= \cM_0 \frac{\alpha}{4 \pi} 
\Big[-\frac{2}{\epsilon_{\mathrm{IR}}^{2}} + \cdots \Big]
\ee
Focusing on the double pole also lets us restrict to just the soft graphs, as they contain the complete soft-collinear singularity.
There are also graphs with one vertex coming from $\Hsc$ and one coming from $H$:
\vspace{-5pt}
 \begin{align}
\cM_C + \cM_D
&= 
\begin{gathered}
\begin{tikzpicture}
 \node at (0,0) {
\parbox{20mm} {
\resizebox{20mm}{!}{
     \fmfframe(0,00)(0,0){
 \begin{fmfgraph*}(40,20)
	\fmftop{R1,R2}
	\fmfbottom{L1}
	\fmf{fermion}{v1,L1}
	\fmf{fermion}{R1,v1}
	\fmf{fermion}{v2,R2}
	\fmf{fermion}{L1,v2}
	\fmf{photon,tension=0}{v1,v2}
        \fmfv{d.sh=circle, d.f=30,d.si=0.1w}{L1}
\end{fmfgraph*}
}}}};
\draw[dashed, line width = 1, darkred] (-0.3,0.5) to (-0.3,-0.7);
\draw[dashed, line width = 1, darkred] (0.6,0.5) to (0.6,-0.7);
\node[above,darkred] at (-0.3,0.4) {${}^{t=-\infty}$};
\node[above,darkred] at (0.6,0.4) {${}^{t=\infty}$};
\end{tikzpicture}
\end{gathered}
+
\begin{gathered}
\begin{tikzpicture}
 \node at (0,0) {
\parbox{20mm} {
\resizebox{20mm}{!}{
     \fmfframe(0,00)(0,0){
 \begin{fmfgraph*}(40,20)
	\fmftop{R1,R2}
	\fmfbottom{L1}
	\fmf{fermion}{v1,L1}
	\fmf{fermion}{R1,v1}
	\fmf{fermion}{v2,R2}
	\fmf{fermion}{L1,v2}
	\fmf{photon,tension=0}{v1,v2}
        \fmfv{d.sh=circle, d.f=30,d.si=0.1w}{L1}
\end{fmfgraph*}
}}}};
\draw[dashed, line width = 1, darkred] (-0.6,0.5) to (-0.6,-0.7);
\draw[dashed, line width = 1, darkred] (0.3,0.5) to (0.3,-0.7);
\node[above,darkred] at (-0.7,0.4) {${}^{t=-\infty}$};
\node[above,darkred] at (0.4,0.4) {${}^{t=\infty}$};
\end{tikzpicture}
\end{gathered}\\
&= \cM_0 \frac{\alpha}{4 \pi} 
\Big[\frac{4}{\epsilon_{\mathrm{IR}}^{2}} +\cdots\Big]
\end{align}
The double IR pole from the $S$-matrix element cancels exactly in the sum $\cM_A + \cM_B + \cM_C + \cM_D$, as anticipated.

 It is worth emphasizing the even the double-pole calculation is not trivial and requires careful manipulation of the distributions involved (cf. Ref.\cite{Frye:2018xjj}). Moreover, the cancellation is different in nature from the cancellation in the computation of a Wilson coefficient. There the soft exchange graph (the analog of $\cM_B$) is {\it subtracted} from $\cM_A$; here the graphs add, with the cancellation coming from graphs $\cM_C + \cM_D$ with one soft and one regular vertex. 

The other TOPT diagrams involving soft-collinear vertices in $\Hsc$, such as
\be
\begin{gathered}
\begin{tikzpicture}
 \node at (0,0) {
\parbox{20mm} {
\resizebox{20mm}{!}{
     \fmfframe(0,00)(0,0){
 \begin{fmfgraph*}(40,20)
	\fmfbottom{B1}
	\fmftop{L1,R1}
	\fmf{phantom}{L1,w1,w2,w3,w4,w5,R1}
	\fmf{phantom,tension=0.5}{B1,x1,v2,w4}
	\fmf{phantom,tension=0.5}{B1,v1,R1}
	\fmffreeze
	\fmf{fermion}{v1,R1}
	\fmf{fermion}{B1,v1}
	\fmf{fermion}{v2,B1}
	\fmf{fermion}{L1,v2}
	\fmf{photon}{v1,v2}
        \fmfv{d.sh=circle, d.f=30,d.si=0.1w}{B1}
\end{fmfgraph*}
}}}};
\draw[dashed, line width = 1, darkred] (-0.3,0.5) to (-0.3,-0.7);
\draw[dashed, line width = 1, darkred] (0.3,0.5) to (0.3,-0.7);
\node[above,darkred] at (-0.4,0.4) {${}^{t=-\infty}$};
\node[above,darkred] at (0.5,0.4) {${}^{t=\infty}$};
\end{tikzpicture}
\end{gathered}
\quad\text{or}\quad
\begin{gathered}
\begin{tikzpicture}
 \node at (0,0) {
\parbox{20mm} {
\resizebox{20mm}{!}{
     \fmfframe(0,00)(0,0){
 \begin{fmfgraph*}(40,20)
	\fmfbottom{B1}
	\fmftop{L1,T1}
	\fmfright{R1}
	\fmf{phantom}{L1,w1,w2,w3,w4,w5,T1}
	\fmf{fermion,tension=0}{L1,w5}
	\fmffreeze
	\fmf{fermion}{w5,B1}	
	\fmf{fermion}{v1,R1}
	\fmf{fermion}{B1,v1}
	\fmf{photon}{v1,w5}
        \fmfv{d.sh=circle, d.f=30,d.si=0.1w}{B1}
\end{fmfgraph*}
}}}};
\draw[dashed, line width = 1, darkred] (-0.3,0.5) to (-0.3,-0.7);
\draw[dashed, line width = 1, darkred] (0.3,0.5) to (0.3,-0.7);
\node[above,darkred] at (-0.4,0.4) {${}^{t=-\infty}$};
\node[above,darkred] at (0.5,0.4) {${}^{t=\infty}$};
\end{tikzpicture}
\end{gathered}
\ee
are not infrared divergent. In fact, the second diagram is zero, because there is no electron-positron annihilation vertex in $\Hsc$. 
Similarly, there are no diagrams with the hard vertex in the asymptotic regions, as $\Hsc$ has only soft and collinear interactions. 

To see the subleading IR poles cancel, we need a regulator other than
pure dimensional regularization, such as offshellness (see~\cite{Manohar:2003vb,Manohar:2006nz}), or explicit phase space restrictions. One should also then include graphs involving the collinear interactions in $\Hsc$ as well as a zero-bin subtraction to avoid overcounting~\cite{Manohar:2006nz}. Using pure dimensional regularization is simplest, since all of the graphs other than $\cM_A$ are scaleless. Thus, after removing UV poles with renormalization, we find
\begin{multline}
\langle e^- | S_H | \gamma^\star  e^-\rangle =  (2\pi)^4\delta^4(q+p_1-p_2)\bar{u}(p_2) \gamma^\mu u(p_1)\\
%\bar{u}(p_1)\gamma^\mu v(p_2) \\ 
\times \left(-i e\right) \Big[ 1 +  \frac{\alpha}{4 \pi} 
\Big(
-\ln ^{2} \frac{\mu^{2}}{Q^{2}}-3 \ln \frac{\mu^{2}}{Q^{2}}-8+\frac{\pi^{2}}{6} \Big)\Big]
\end{multline}
To confirm that the IR divergences cancel in $S_H$, without invoking scaleless-integral magic, we can impose physical cutoffs on the degrees of freedom that interact in $\Hsc$, such as including only photons with energy less than $\delta$ or within angle $R$ of an electron~\cite{Sterman:1977wj}. Then the diagrams like $\cM_B$
are no longer scaleless. We have checked that all of the IR divergences cancel in $S_H$ using this approach. Although $S_H$ comes out IR finite, it retains sensitivity to the scales $R$ and $\delta$; in pure dimensional regularization, these cutoff scales are replaced by the single scale $\mu$. 

With a new definition of the $S$-matrix, it is natural to ask what are its predictions for observables? Consider an infrared-finite observable, such as the total cross
section in $Z\to$ hadrons. To compute it, note that the total cross section for $Z \to $ {\it anything}, at order $\alpha_s$ is zero, since the forward scattering $Z \to Z$ cross section exactly cancels the cross section to everything else. This follows from unitarity, whether using $S_H$ or $S$. Now, the $Z$ has no soft or collinear interactions, so $|Z^d\rangle = |Z\rangle$. Thus $\langle Z | S_H | Z \rangle =  \langle Z | S | Z \rangle$ to all orders in perturbation theory. Therefore the $Z\to Z$ forward-scattering cross section is the same with $S_H$ and $S$, and so is the  $Z\to$ hadrons cross section.

% If we break down the contributions, we see that with $S_H$, the matrix elements for $Z\to q\bar{q}$ and $Z \to q \bar{q} g$ are separately infrared finite. Moreover, the  $Z \to q \bar{q} g$ matrix element naturally vanishes in soft or collinear limits, as in these limits the amplitude for emitting a gluon using $H$ is canceled by the amplitude for emitting a gluon in the asymptotic region using $\Hsc$. This is analogous to how the matching coefficient for a 3-jet operator is computed in SCET (by subtracting the previously-matched 2-jet operator contribution~\cite{Bauer:2006qp,Bauer:2006mk}). 
% Thus not only are $S_H$ amplitudes IR finite, but with $S_H$ the contributions to the cross section for $Z\to$ hadrons from a $q\bar{q}$ state and $q\bar{q} g$ states are separately infrared finite. 

More generally, if we consider an observable less inclusive than the total cross section, such as a jet rate, then the details of the asymptotic dynamics will be important to determining the differential cross section. When we include this dynamics by evolving the final state with an $e^{-i \Has t_+}$ factor, we would effectively be computing
$\sum_X |\langle X | e^{-i \Has t} S_H |Z\rangle|^2 = \sum_X |\langle X | S |Z\rangle|^2$, so
the differential cross section will agree {\it exactly} with one computed using $S$. Since infrared-safe cross sections computed using $S$ are incontrovertible agreement with data, this is reassuring: we have not created more problems than we have solved with a finite $S$-matrix. On the other hand, there are also issues where physical predictions using $S$ are ambiguous, such as with charged particles in the initial states. $S_H$ could possibly shed light on these processes.

%Unfortunately
%Although $\cN  =4$ is conformal, its $S$-matrix is still IR divergent.  
%One approach to handling the divergences in $S$ is simply to drop the $\frac{1}{\epsilon_\text{IR}}$ terms. 
%Instead, one can subtract not just the IR poles, but also some finite parts form 
%,Golden:2019kks,Caron-Huot:2019vjl}.
Having a finite $S$-matrix is perhaps most appealing in situations where the $S$-matrix is of interest for its own sake, for example, for its mathematical properties. One popular playground for studying the mathematics of the $S$-matrix is $\cN=4$ super-Yang-Mills theory. This theory is a conformal gauge field theory. Although its $S$-matrix is UV finite, it is still IR-divergent. Moreover, its mathematical properties depend on how these IR divergences are removed. For example, 
the simplest approach is simply to drop the $\frac{1}{\epsilon_\text{IR}}$ terms, $\msbar$-style. 
Doing so for the planar 2-loop 6 particle amplitude, for example, gives
a complicated function of the 9 kinematical invariants. If instead one employs the BDS-Ansatz, taking the ratio of 
the $S$-matrix element to the exponentiation of the 1-loop result~\cite{Bern:2005iz,Anastasiou:2003kj}, then the result is a relatively simple ``remainder function" of only the three dual-conformally invariant cross-ratios~\cite{DelDuca:2010zg,Goncharov:2010jf}. 
While dual-conformal invariance is preserved by the BDS-Ansatz, the BDS remainder functions have unappealing analytic properties, such as violating the Steinmann relations~\cite{Caron-Huot:2016owq}. A BDS-like Ansatz might preserve these~\cite{Alday:2009dv}. A minimal normalization
is another option~\cite{Golden:2019kks}. In the computation of $S_H$, the IR divergences cancel automatically:
the analog of the BDS subtraction comes naturally from multiplying $1/\epsilon$ counterterms for $S_H$ with the finite $\cO(\epsilon)$ parts $S_H$-matrix elements. Thus $S_H$-matrix elements provide some of the benefits of IR-finite remainder functions, without the arbitrariness of a ratio. Moreover, as the $S_H$ operator is unitary, properties that follow from unitarity (perhaps including the Steinmann relations) should be automatically satisfied. This is in contrast to remainder functions which are quotients of $S$-matrix elements to other quantities.

% In this way, a remainder function can be interpreted as an $S_H$-matrix element, rather than requiring an IR-universality inspired~\cite{Gardi:2009qi,Becher:2009qa}
%  but ad hoc subtraction.  
% Dual conformal invariance can be understood in part from a mapping between UV-finite planar amplitudes and IR-finite rescaling-invariant Wilson loops.
% In a sense, $S_H$-matrix elements provide a much broader class of IR-finite quantities than Wilson loop expectation values,
% and may exhibit symmetries like dual-conformal invariance that are obscured by the IR regulator required for $S$-matrix computations. 

In this paper, we have argued that there is nothing sacred about the traditional $S$-matrix. Its non-perturbative definition is absurdly complicated, and its interaction-picture definition involves an admixture of free and full-theory time evolution.
In a theory with massless particles, it is natural to replace the free evolution with universal soft and collinear evolution.
Unlike $S$, whose matrix elements are either infinite (IR divergent) or zero (after exponentiation of the IR divergences),
matrix elements of this new object $S_H$ are IR finite to all orders.

In summary, this paper provides the first explicit construction of a $S$-matrix for non-Abelian gauge theories with no collinear or soft divergences; it provides rules (see also~\cite{Hannesdottir:2019opa}) for computing $S_H$ beyond just the cancellation of the singularities, allowing the mathematical properties of the $S$-matrix to be explored with the IR-divergence problem removed in a natural way; it connects to previous literature on dressed/coherent states, but also argues that such non-normalizable states are not needed for $S_H$ or to compute observables; finally, it connects $S_H$-matrix elements to SCET and to remainder functions in $\cN=4$ SYM theory for the first time. While there is much still to be understood about $S_H$, it provides a solid starting point for an improved understanding of scattering in theories with massless particles.

%
%The main point of this paper is that there is nothing sacred about the traditional $S$-matrix. The $S$-matrix is {\it not} the projection of  at $t=-\infty$ 
%
%
%While $S_H$ may maintain interesting properties. 
%
%In addition to its symmetries, there is much else to be understood about $S_H$.  Unlike $S$-matrix elements which are either infinite (IR divergent) or zero (after exponentiation of the IR divergences), matrix elements for $S_H$ are finite to all orders. Nevertheless, they will be asymptotic series in the couplings and new approaches are needed to study their non-perturbative structure. It could also be interesting to study the UV structure of $S_H$, its scheme sensitivity, and its properties in gravitational and conformal field theories. 

We would like to thank S. Caron-Huot and J. Collins for invaluable discussions. This work was supported in part by the U.S. Department of Energy under contract DE-SC0013607. 
\end{fmffile}

\bibliography{FiniteS.bib}

\end{document}